\documentclass[reprint,superscriptaddress,nobibnotes,amsmath,amssymb,aip]{revtex4-2}
\usepackage{showyourwork}
\usepackage[version=4]{mhchem}
\usepackage{graphicx}
\usepackage[utf8]{inputenc}
\graphicspath{{figures/}}
\usepackage{bm}
\usepackage{setspace}
\usepackage{siunitx}
\usepackage{lipsum}
\newcommand{\papertitle}{On the Estimation of Centre of Mass in Periodic Systems}
\usepackage[noabbrev,nameinlink,capitalize]{cleveref}
\crefname{equation}{Eq.}{Eqs.}
\crefformat{equation}{Eq.~#2#1#3}
\crefrangeformat{equation}{Eqs.~#3#1#4 to~#5#2#6}
\crefmultiformat{equation}{Eqs.~#2#1#3}{ and~#2#1#3}{, #2#1#3}{ and~#2#1#3}
\crefname{figure}{Fig.}{Figs.}

\makeatletter
\def\@email#1#2{%
 \endgroup
 \patchcmd{\titleblock@produce}
  {\frontmatter@RRAPformat}
  {\frontmatter@RRAPformat{\produce@RRAP{*#1\href{mailto:#2}{#2}}}\frontmatter@RRAPformat}
  {}{}
}%
\makeatother

\begin{document}

\let\oldaddcontentsline\addcontentsline
\renewcommand{\addcontentsline}[3]{}

\title[Accurate Centre of Mass Estimation]{\papertitle}

\author{Harry Richardson}
  \affiliation{Centre for Computational Chemistry, School of Chemistry, University of Bristol, Cantock's Close, Bristol, BS8 1TS, UK.}
\author{Josh Dunn}
  \affiliation{Centre for Computational Chemistry, School of Chemistry, University of Bristol, Cantock's Close, Bristol, BS8 1TS, UK.}
\author{Andrew R. McCluskey}
  \affiliation{Centre for Computational Chemistry, School of Chemistry, University of Bristol, Cantock's Close, Bristol, BS8 1TS, UK.}
  \affiliation{Diamond Light Source, Harwell Campus, Didcot, OX11 0DE, UK.}
\email{josh.dunn@bristol.ac.uk and andrew.mccluskey@bristol.ac.uk}

\date{\today}

\begin{abstract}
Calculation of the centre of mass of a group of particles in a periodically-repeating cell is an important aspect of chemical and physical simulation. 
One popular approach calculates the centre of mass via the projection of the individual particles' coordinates onto a circle [Bai \& Breen, \emph{J. Graph. Tools}, \textbf{13}(4), 53, (2008)].
However, this approach involves averaging of the particles in a non-physically meaningful way resulting in inaccurate centres of mass. 
Instead the intrinsic weighted average should be computed, but the analytical calculation of this is computationally expensive and complex. 
Here, we propose a more computationally efficient approach to compute the intrinsic mean that is suitable for the majority of chemical systems. 
\end{abstract}

\maketitle

\section{Introduction}
\label{sec:intro}

\begin{figure*}
  \includegraphics[width=\textwidth]{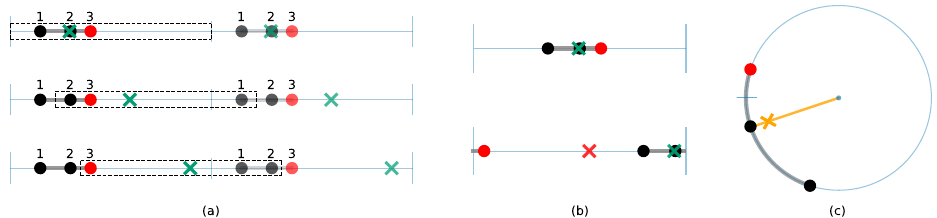}
  \caption{
  (a) Demonstration of the $N$ possible centres of mass in an $N$ particle system, where two periodic cells are visible, and the dashed line indicates the particle grouping that leads to a given centre of mass (green cross). 
  (b) The problem with the naïve centre of mass (red cross) calculation in periodic systems; where the top shows the particle group in a single cell, compared with the bottom where the particle group spans a periodic boundary. 
  (c) The projection of the particles in the bottom diagram of (b) onto a circle to find the centre of mass (yellow cross), which minimises the weighted distance to all particles on the planar disk of the circle.}
  \label{fig:problems}
  \script{Fig1.py}
\end{figure*}

The calculation of the centre of mass of a group of particles, such as in a molecule, is important across computational chemistry and physics.\cite{zhang_chemically_2024,happel_coordinated_2024,maggi_universality_2021,grillo_molecular_2023,bullerjahn_unwrapping_2023,jaeger-honz_systematic_2024}
However, computing the centre of mass is non-trivial for systems with periodic boundary conditions. 
Any group of particles will have the same number of valid centres of mass---positions in the simulation cell where the weighted relative positions of the masses sum to zero---as the number of particles (\cref{fig:problems}a). 
Computing the centre of mass will typically form part of some analysis that is performed after the simulation has been run. 
Therefore, the simulation trajectory may involve large numbers of molecules or long trajectories, leading to a large number of individual centre of mass calculations. As such, efficient computation and low computational overhead is of high importance. 

In the physical sciences, we may be concerned with the centre of mass of a single bonded molecule or a defined group of particles.
For such systems, there is only one correct answer for the centre of mass.
Practically speaking, for most simulations, it is desirable to compute the centre of mass that minimises the squared periodic distance to the measured particles (Euclidean or \( L^2 \) norm).

In a non-periodic system, identifying the ``correct'' centre of mass of a group of particles, the average position weighted by the masses, is trivial. 
However, where the periodic boundary can intersect the particle group, this naïve calculation may be ``incorrect'' (\cref{fig:problems}b). 
This problem can be resolved where bonding or other relational information is present, but this is not always the case for computational simulation trajectories. 

Bai and Breen proposed an algorithm to efficiently estimate the centre of mass,\cite{bai_calculating_2008} involving the projection of each orthogonal Cartesian dimension onto a circle to find the centre of mass in the periodic space (\cref{fig:problems}c). 
However, this approach appears to introduce numerical error when compared to the continuum centre of mass, regardless of whether particles span the periodic boundary. 
For the original use case, to allow the recentering and visualisation of groups of aggregated particles, the method is effective, as numerical accuracy is less important. 
However, within the context of computational chemistry and physics, this is not the case.  
Here, we discuss the nature of this error, present a general solution, and an optimised approach that is accurate for the majority of chemical systems.

\section{Circular Weighted Average}

To outline the circular averaging approach from Bai and Breen, known in directional statistics as the extrinsic mean,\cite{hotz_intrinsic_2014} we consider a simple one-dimensional periodic system, where the cell ranges from \num{0} to \num{1}. 
This system is analogous to a fractional coordinate system. 
A group of $N$ particles can be described with two vectors, one for the positions, $\bm{x}$, and another for the masses, $\bm{m}$. 
Each particle $i$ is then projected onto the two dimensions of a unit circle, $\bm{\xi}$ and $\bm{\zeta}$, where, 
\begin{equation}
    \xi_i = \cos(2\pi x_i),
\end{equation}
and 
\begin{equation}
    \zeta_i = \sin(2\pi x_i).
\end{equation}
The average of the vectors $\bm{\xi}$ and $\bm{\zeta}$ are then found, weighted by the masses, $m$, of $i$,
\begin{equation}
    \bar{\xi} = \frac{\sum_{i=1}^{N}m_i\xi_i}{\sum_{i=1}^{N}m_i},
\end{equation}
and
\begin{equation}
    \bar{\zeta} = \frac{\sum_{i=1}^{N}m_i\zeta_i}{\sum_{i=1}^{N}m_i}.
\end{equation}
$\bar{\xi}$ and $\bar{\zeta}$ represent the coordinates of a point on the plane formed by the circle. 
The projection process is then reversed for the weighted average circular coordinates.
Practically, this is achieved using the 2-argument arctangent, 
\begin{equation}
    \bar{x} = \frac{\text{atan}2(-\bar{\zeta}, -\bar{\xi}) + \pi}{2\pi},
\end{equation}
to give the centre of mass position.
In the case of an orthogonal multi-dimensional system, this method can be repeated independently for each dimension. 

The approach outlined above provides the average position in a two-dimensional coordinate system, and then projects this onto a position on the circle corresponding to a position in periodic space (\cref{fig:problems}c). 
However, this projection is from a position with no physical meaning in the context of the periodic cell, i.e., the position does not sit on the manifold of the circle itself.
The result of using a position on the plane of the circle, rather than the circle itself, is a discrepancy when comparing the circular weighted average centre of mass with that of the system in a non-periodic space. 

\begin{figure}
    \centering
    \includegraphics[width=8cm]{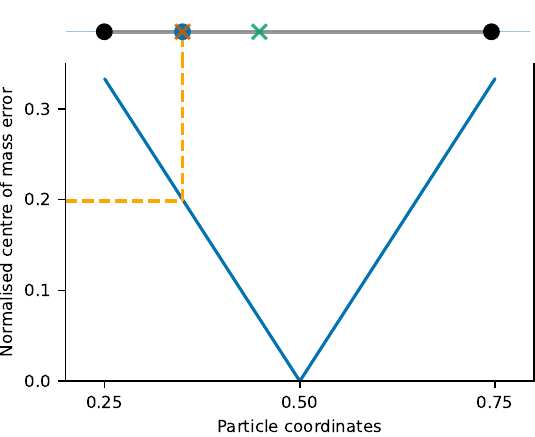}
    \caption{Quantification of the limiting case (bottom), a 3-particle group spanning close to half the periodic box, where the solid blue line shows the error between the circular weighted average centre of mass and continuum centre of mass normalised by the span (length) of the particle group.
    This error is shown visually for a single example (top), where the circular averaged centre of mass (orange cross) is not close to the true centre of mass (green cross).}
    \label{fig:error_quantification}
    \script{Fig2.py}
\end{figure}

Consider the limiting case of a 3-particle system, where a central particle is moved between 2 edge particles, which are stationary at coordinates of 0.26 and 0.74. 
This group of particles does not span the periodic cell and therefore the continuum centre of mass can be computed naïvely as the weighted average of the positions. 
\cref{fig:error_quantification} presents the difference between the naïve centre of mass and that computed with the circular weighted average approach described above, normalised by the total extent of the group of particles. 
The error in the circular weighted average varies, reaching a maximum of more than \SI{30}{\percent} of the particle group size when the central particle is very close to the stationary particles, where the system is highly asymmetric. 
At the point where the particles are equally distributed, the system is perfectly symmetric, and there is no error in the centre of mass. 
Similarly, there is no error in one or two particle systems as these are inherently symmetrical (see Appendices for further discussion of the importance of asymmetry). 

\section{Intrinsic Weighted Average}

The mean of a set of samples is the value that minimises the sum of the squared distances to the samples.\cite{neilsen_matrix_2012}
The circular weighted average centre of mass satisfies this criterion by minimising the chordal distance between the points on the planar disk formed by the circle.\cite{hotz_non-asymptotic_2016}
However as outlined above, this position is not physically meaningful and requires projection onto the circular manifold, leading to the inconsistency with the continuum centre of mass.

The position that matches the continuum mean of a periodic system is the intrinsic mean, which minimizes the sum of the squared arc distances between the mean and the samples. 
As all the points exist on the periodic unit circle, all positional information is contained in their angular position. 
This results in a computationally expensive minimisation problem,\cite{hotz_extrinsic_2013} for which an analytical solution for the intrinsic mean can be found by producing $N$ candidate means,
\begin{equation}
\bar{x}_{n} = \bar{x}_0 + \frac{2\pi n}{N}, \quad n = 1, 2, \dots, N,
\end{equation}
where $\bar{x}_0$ represents a naïve mass-weighted average position for all particles in the group.
The true intrinsic mean is then the value in the vector, $\bar{\bm{x}} = [\bar{x}_{1}, \bar{x}_{2}, \ldots \bar{x}_{n}]$, with the lowest sum of squared arc distances between the points.
The method above will always produce the true intrinsic mean and, therefore, the correct continuum centre of mass. 
However, this approach has a high computational complexity and even an optimised approach is still computationally expensive with $\mathcal{O}(N\log N)$ scaling due to the need to sort the particle positions.~\cite{hotz_extrinsic_2013,miolane_geomstats_2020}

It should be noted that for bonded molecules that span over half the periodic box, the intrinsic and molecular centre of masses are not necessarily the same.
The molecular centre of mass must be within the molecule even if this is not the minimum of the squared arc distances within the system.
Therefore, for such systems, bonding information is required to calculate the molecular centre of mass. 

\section{\emph{Pseudo}-Centre of Mass Recentering}

We propose a computationally efficient approach, that finds the true continuum centre of mass and is significantly faster than the optimised intrinsic mass computation above (see Appendix~\ref{app:complex}). 
This approach involves using the circular weighted averaged centre of mass as a \emph{pseudo}-centre of mass. 
The group of particles is repositioned to the middle of the simulation cell using the \emph{pseudo}-centre of mass, such that there is no risk that the group spans a periodic boundary condition. 
Then, the standard weighted average may be used to find the centre of mass before restoring this to the original space. 

The use of this recentering assumes that the circular weighted average centre of mass method finds a value close enough to the true centre of mass that the recentering will place all particles contiguously in the periodic cell. 
This assumption holds true for all systems where the particle group is less than half the extent of the periodic cell and is often still true unless the system is very asymmetric. 
If a group of particles spans more than half of the periodic box, the analytic intrinsic weighted average approach should be used. 

\begin{figure}
    \centering
    \includegraphics[width=8cm]{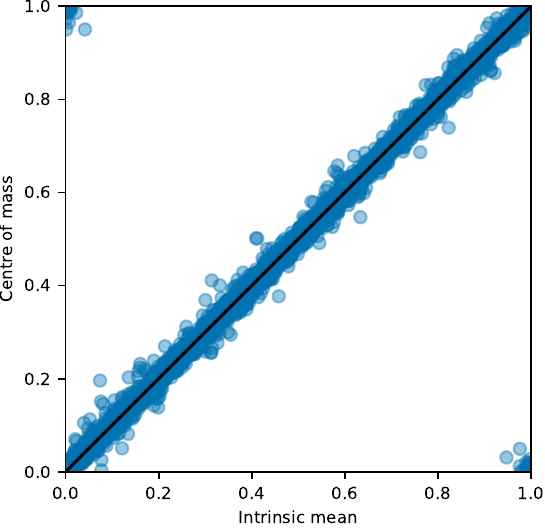}
    \caption{A comparison of the circular weighted averaged centre of mass method (spread of blue dots) and \emph{pseudo}-centre of mass recentering approach (black line) with the intrinsic weighted average centre of mass. 
    $2^{18}$ random configurations of one-dimensional periodic boundary-spanning particle groups were generated, with between 3 and 512 particles and a total length of less than half the box. 
    The apparent outliers in the top-left and bottom-right corners are not true outliers, rather, they arise when the intrinsic mean approaches the periodic boundary, the error in the circular weighted averaging centre of mass causes the estimate to wrap onto the opposite side of the periodic cell.
    }
    \label{fig:method_valid}
    \script{Fig3.py}
\end{figure}

To demonstrate the effectiveness of our \emph{pseudo}-centre of mass recentering approach, \cref{fig:method_valid} compares it to the circular weighted average centre of mass. 
As can be seen the pseudo-centre of mass recentering approach produces a accurate estimate of the intrinsic center of mass, provided that particles do not span more half the box. 
Therefore, for most chemical simulations, this \emph{pseudo}-center of mass recentering approach should be used to ensure accuracy in the centre of mass calculation. 

\section{Conclusions}

We have discussed the nature of centre of mass computation for periodic chemical simulations.
We highlighted the problems of computing the circular weighted average centre of mass versus the intrinsic weighted average centre of mass, indicating that the latter matches the continuum centre of mass that is expected. 
However, computation of intrinsic weighted average centre of mass is computationally complex and may be inefficient for the analysis of large trajectories. 
To address this, we propose an efficient approach based on a \emph{pseudo}-centre of mass recentering that finds the intrinsic centre of mass in the majority of chemical systems. 
To enable the use of this approach, we have implemented it in the open-source Python package \textsc{kinisi}\cite{mccluskey_kinisi_2024} and hope that others will consider its use in future. 

\begin{acknowledgements}
We thank Richard Gowers for the insightful discussion that started us down this rabbit hole and for helpful comments on the manuscript.
We also thank Neil L. Allan for helpful discussions that lead to this work. 
We acknowledge Karol Ławniczak's insightful lectures on directional statistics that helped us to better understand this subject.
H.R. acknowledges the Engineering and Physical Sciences Research Council for DTP funding (EP/W524414/1). 
BP plc is thanked for the financial support of J.D. through an industrial CASE (ICASE) studentship in partnership with the Engineering and Physical Sciences Research Council (EP/T51763X/1).
\end{acknowledgements}

\section*{Conflict of Interest Statement}

The authors have no conflicts to disclose.

\section*{Author Contributions}
H.R.: Formal analysis, Investigation, Methodology, Software, Visualisation, Writing – original draft.
J.D.: Conceptualization, Investigation, Methodology, Supervision, Writing – review \& editing.
A.R.M.: Formal analysis, Funding acquisition, Investigation, Methodology, Project administration, Supervision, Writing – review \& editing.

\section*{Data Availability}

The data that support the findings of this study are openly available in a GitHub repository,\cite{richardson_github_2025} under an MIT license, including a complete set of analysis/plotting scripts allowing for a fully reproducible and automated analysis workflow, using \textsc{showyourwork}.\cite{luger_showyourwork_2021}

\appendix

\section{Mathematical Equivalency of Polar Averaging and First Moment Methods}
\label{app:math}

This appendix presents the mathematical equivalency between the circular weighted average approach to compute the centre of mass, and finding the maximum of the first term of the Fourier series of the mass density as a function of position, known as the first moment method.\cite{teague_robust_2018}

For a real-valued function, the Fourier series can be described as 
\begin{equation}
    s_N(x) = a_0 + \sum_{n=1}^{N} \left[ a_n \cos \left( 2\pi \frac{n}{P} x \right) + b_n \sin \left( 2\pi \frac{n}{P} x \right) \right],
    \label{equ:fseries}
\end{equation}
where $N$ is some finite number of terms that are computed, and $P$ is the period of the function. 
Eqn.~\ref{equ:fseries} is equivalent to
\begin{equation}
    s_N(x) = A_0 + \sum_{n=1}^{N} A_n \cos\left(2\pi \frac{n}{P}x - \varphi_n\right),
\end{equation}
by the identity, 
\begin{equation}
    \begin{aligned}
        \cos \left( 2\pi \frac{n}{P} x - \varphi_n \right) = & \cos(\varphi_n) \cos \left( 2\pi \frac{n}{P} x \right)  \\ 
        & + \sin(\varphi_n) \sin \left( 2\pi \frac{n}{P} x \right),
    \end{aligned}
\end{equation}
where $\varphi_n$ is the phase shift of the $n$-th harmonic. 
If we write, 
\begin{equation}
    a_n = A_n \cos(\varphi_n),
\end{equation}
and
\begin{equation}
    b_n = A_n \sin(\varphi_n).
\end{equation}
Then $a_n$ and $b_n$ can be considered to be constants that ``weight'' the relative contributions of $\sin \left( 2\pi \dfrac{n}{P} x \right)$ and $\cos \left( 2\pi \dfrac{n}{P} x \right)$, akin to the weighted average of $\bm{\xi}$ and $\bm{\zeta}$ carried out within the circular weighted average approach.

As stated in the main text, the circular weighted average approach is equivalent to finding the maxima of the first function of the Fourier series, $f_1$,
\begin{equation}
    f_1(x) = a_1 \cos{\Big(2\pi\frac{1}{P} x\Big)} + b_1 \sin{\Big(2\pi\frac{1}{P} x\Big)},
\end{equation} 
where $P$ is $1$ as the circular weighted average approach projects the full box size onto a single unit circle. 
To maximise the value of $f_1(x)$ a stationary point is found where,
\begin{equation}
    \frac{\partial f_1(x)}{\partial x} = 0,
\end{equation}
which is 
\begin{equation}
    \frac{\partial f_1}{\partial x} = -2 \pi a_1 \sin( 2\pi x) + 2 \pi b_1 \cos( 2\pi x) = 0.
    \label{equ:der}
\end{equation}
Eqn.~\ref{equ:der} can be rearranged to give, 
\begin{equation}
\frac{\sin(2 \pi x)}{\cos(2 \pi x)} =  \frac{b_1}{a_1}
\end{equation}
and from the trigonometric identity, $\tan(x) = \sin(x) / \cos(x)$, 
\begin{equation}
    \tan(2 \pi x) = \frac{b_1}{a_1}.
\end{equation}
Therefore the position at which the first function of the Fourier series is maximised, $x$, can be calculated as, 
\begin{equation}
    x = \frac{\text{atan}2(b_1, a_1)}{2\pi}.
\end{equation}
In terms of periodic position, 
\begin{equation}
    \frac{\text{atan}2(b_1, a_1)}{2\pi} \equiv \frac{\text{atan}2(-b_1, -a_1) + \pi}{2\pi}.
\end{equation}
The negative $b_1$ and $a_1$ in the circular weighted average approach ensure that the centre of mass is always in the correct periodic box with no additional wrapping.

\section{Behaviour of Extrinsic Mean With Asymmetry}
\label{app:fourier}

\begin{figure}
    \centering
    \includegraphics[width=8cm]{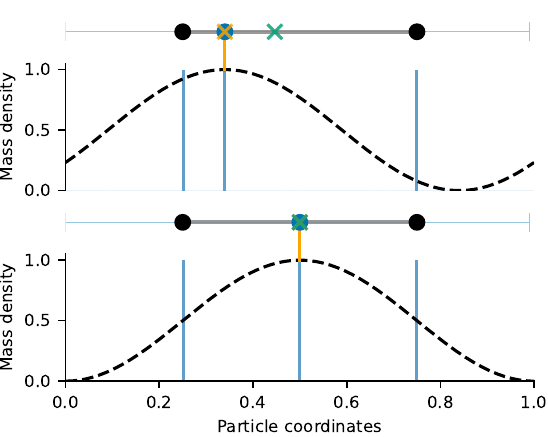}
    \caption{Mass density histograms for two 3 particle molecules, (top) asymmetrically distributed particles [0.26, 0.34, 0.74], (bottom) symmetric molecule [0.26, 0.50, 0.74] (blue bars), and the sine waves produced from the first components of the fourier series (dashed lines). It can be seen in the top image that the asymmetry in the molecule causes the maximum of the sine wave to be a poor estimator of the intrinsic centre of mass. }
    \label{fig:sine_wave_assym}
    \script{Fig4.py}
\end{figure}

As described in Appendix~\ref{app:math}, the circular weighted average centre of mass can be interpreted as identifying the maximum of the first term in the Fourier series, the first moment, of the mass density distribution. 
The first moment is the the peak position of a sine wave with a period of the periodic box length. 
\cref{fig:sine_wave_assym} shows mass density profiles and corresponding first moment sine waves for two three-particle systems, where it can be seen that the sine wave peak tracks the position of the moving particle. 
However, this does not accurately represent the intrinsic center of mass, except in the special case where the particles are symmetrically distributed. 

In \cref{fig:sine_wave_assym}, we show this sine wave for a system with particles at 0.26,  0.34, 0.74, where it is clear that a single sine wave is not capable of describing this asymmetric function. 
This is compared with a system where the particles are symmetrically positioned, in which the circular weighted averaging approach accurately estimates the continuum centre of mass. 
By describing an asymmetric function with a single sine wave, we produce an incorrect estimate of the intrinsic centre of mass. 
In this system, the extrinsic centre of mass will always be the position of the central particle, as the outer particles are almost exactly opposite each other on the circular manifold.
Therefore, when projected onto the circular plane, the extrinsic centre of mass of the outer particles cancel each other out, leaving the total extrinsic centre of mass to be determined by the central particle. 
First-moment-based methods such as this are known to be sensitive to asymmetry~\cite{teague_robust_2018} as a single sine wave cannot encode asymmetrical information effectively. 

\begin{figure}
    \centering
    \includegraphics[width=8cm]{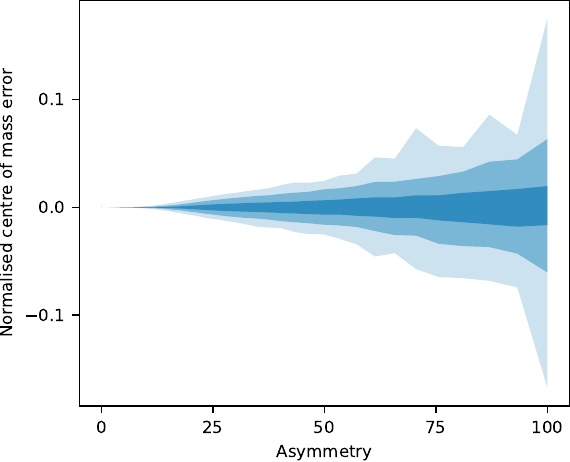}
    \caption{Error in circular averaged centre of mass method (blue shading representing 1, 2, and 3$\sigma$ spreads) as a function of asymmetry.
    $2^{18}$ random configurations of one-dimensional periodic boundary-spanning particle groups were generated (the same samples as in \cref{fig:method_valid}, with between 3 and 512 particles and a total length of less than half the box). 
    For each configuration, the two approaches were used, and the asymmetry of the particle group~\cite{xioajun_on_1991} is compared to the error in the estimate normalised by the span (length) of the particle group. 
    The results were binned along the $x$-axis using 100 bins spaced evenly on a log scale.}
    \label{fig:method_comparison}
    \script{Fig5.py}
\end{figure}

To validate and visualise the effect of asymmetry on the deviation of the circular weighted averaging centre of mass from the continuum centre of mass, the error—defined as the absolute difference between the intrinsic centre of mass and the circular averaged centre of mass, normalised by the span of the particle group was calculated.
This error is compared with the asymmetry of the particle mass density.\cite{xioajun_on_1991}
In \cref{fig:method_comparison}, as the asymmetry increases, so does the difference between the circular averaged centre of mass and the intrinsic centre of mass. 

\section{Comparison of Computational Cost}
\label{app:complex}

The computational cost of centre of mass calculation is an important factor to allow for efficient simulation analysis. 
Large simulations may involve many thousands or millions of particles over as many frames~\cite{lu_pflops_2021,stevens_molecular_2023} resulting in a very large number of calculations of the centre of mass of a particle group.
As such, it is valuable to compare the computational efficiency of the two algorithms. 

With regards to computational scaling, the rate limiting step of the intrinsic centre of mass calculation, as implemented in \textsc{geomstats},\cite{miolane_geomstats_2020} is the sorting of the particle positions. 
Sorting algorithms are well documented to scale, at best, as $\mathcal{O}[N log(N)]$,\cite{knuth_sorting_1998} where $N$ is the number of particles in the group.
Meanwhile, all steps in our \emph{pseudo}-centre of mass recentering algorithm scale as $\mathcal{O}(N)$. 

\begin{figure}
    \centering
    \includegraphics[width=8cm]{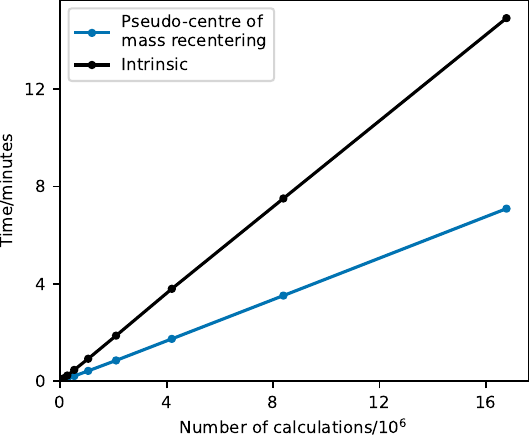}
    \caption{Increasing number of calculations of the centre of mass of an example 20 particle system against computational time for both methods. 
    All calculations run on an Apple Macbook Pro with M3 chipset.}
    \label{fig:comp_efficiency}
    \script{Fig6.py}
\end{figure}

However, in reality, computational scaling with the number of particles is not of great importance, as typical analyses would involve many small calculations of the centre of mass, such that constant computational overheads will dominate. 
To compare computational efficiencies, \cref{fig:comp_efficiency} presents the time taken for both algorithms with increasing number of iterations for a \num{20} particle systems.
The \emph{pseudo}-centre of mass recentering method is more than twice as fast as the \textsc{geomstats} implementation of the intrinsic method, due to reduced complexity and number of steps. 

\bibliography{bib}

\end{document}